\documentclass[12pt,dvips]{article}

\usepackage{amsmath,amssymb,exscale}
\usepackage{array,multicol}
\usepackage{afterpage,float,flafter}
\usepackage{epsfig,rotating,pifont,fancybox}
\usepackage{cite}
\makeatletter
\@addtoreset{equation}{section}
\makeatother

\title{
\vspace{1cm} 
\Large\textbf{
Grand Unification in RS1
}
\vspace*{.5cm}
\author{\large \textbf{
Kaustubh Agashe\footnote{email: kagashe@pha.jhu.edu} , 
Antonio Delgado\footnote{email: adelgado@pha.jhu.edu} 
\mbox{  }and Raman Sundrum\footnote{email: sundrum@pha.jhu.edu}}\\
\emph{
Department of Physics and Astronomy} \\ 
\emph{Johns Hopkins University} \\ 
\emph{3400 North Charles St}. \\ 
\emph{Baltimore, MD 21218-2686}}}

\date{}
\begin{document}
\maketitle
\thispagestyle{empty}
\vspace*{.5cm}
  
\begin{abstract} 
We study unification in the  
Randall-Sundrum scenario for solving the hierarchy problem, with gauge fields
and fermions in the bulk. 
We calculate the one-loop
corrected low-energy effective gauge
couplings in a unified theory, broken at the 
scale $M_{GUT}$ in the bulk. We find that, 
although this scenario has an extra dimension, there is a robust
(calculable in the effective field theory) 
logarithmic dependence on $M_{GUT}$, strongly suggestive of {\em high}-scale
unification, very much as in  
the ($4D$) Standard Model. 
Moreover,
bulk threshold effects are naturally small,
but volume-enhanced, 
so that
we can accommodate the measured gauge couplings. 
We show in detail how excessive proton decay is forbidden by an extra
$U(1)$ bulk {\em gauge} symmetry. 
This mechanism requires us to further break the
unified group using boundary
conditions.
A $4D$ dual interpretation, 
in the sense of the AdS/CFT correspondence, is provided for all
our results.
Our results show that
an attractive unification mechanism can 
combine with a {\em non}-supersymmetric
solution to the hierarchy problem.

\end{abstract} 
  
\newpage 
\renewcommand{\thepage}{\arabic{page}} 
\setcounter{page}{1} 
  
\section{Introduction} 
Grand unification
ideas \cite{gut} offer an attractive theoretical framework for physics
beyond the standard model (SM). 
Phenomenologically, unification helps to explain the gauge
quantum number assignments of the SM,
while correlating the observed
gauge couplings. It also seems significant for ultimate gauge-gravity
unification that the high unification scale
is not too far from the Planck scale. This also helps to suppress baryon decay.
As is well known, unification has achieved its most
striking success within the scenario of weak scale supersymmetry (SUSY)
(for a review, see reference \cite{raby}). 
SUSY 
protects the electroweak scale from radiative corrections at
the unification and Planck scales, while supplying just the right complement of
superpartners to give a fairly precise meeting of gauge couplings as 
they are run up in
energy. Only believably small threshold corrections need to  be invoked. 

In this paper, 
we will study unification in quite a different, {\em non}-supersymmetric,
approach to the hierarchy problem, namely the Randall-Sundrum (RS1) 
extra-dimensional
scenario \cite{rs1}. Here, the hierarchy problem is 
solved by having a highly warped
compactification with the metric:
\begin{eqnarray} 
ds^2 & = & e^{-2k | y | } \eta_{\mu \nu} dx^{\mu} dx^{\nu} + dy ^2, 
\; - \pi r_c \leq y \leq \pi r_c,
\label{metric}
\end{eqnarray} 
a slice of AdS$_5$ with radius of curvature $k^{-1}$. The extra dimension is an
orbifolded circle of 
``radius'' 
$r_c$.
In the RS1 set-up, the relationship between $5D$ mass scales 
(which are all taken to be of order the $4D$ Planck scale, $M_{Pl}$) and those in
an effective $4D$ description depends on the location in the extra dimension:
\begin{equation}
M_{4D} \sim M_{5D} e^{-k y}.  
\end{equation}
Thus,  
the Higgs sector is localized at the ``IR brane''
($y = \pi r_c$), where it is protected by a low
warped-down fundamental scale of order a TeV
for a modestly large radius, while $4D$ gravity is localized near the 
``Planck
brane'' ($y = 0$) and has a Planckian fundamental scale \cite{rs2}.
If the 
SM (gauge) fields are on the IR brane as in the 
original RS1 model \cite{rs1}, then
the effective UV cut-off for the SM is $\sim$ TeV and 
hence unification of gauge
couplings at Planckian scales cannot be addressed in the
effective field theory. 
If the SM gauge fields are in the bulk (first suggested
in reference 
\cite{wise} and studied in references 
\cite{hewettpomarol}), then quantum loops are sensitive to
greatly varying, in particular, Planckian scales since the loops 
span the extra dimension. 
Hence high scale unification can be studied.
We will give realistic models based on having unified $SU(5)$ gauge fields and
SM fermions propagating
in the extra-dimensional ``bulk''. 

Our models display several attractive features: (i) They
help to explain the observed gauge quantum numbers of quarks 
and leptons by greatly
restricting the possibilities. 
(ii) They correlate the observed gauge couplings by a
mechanism of logarithmic gauge coupling unification. 
The quantitative success is very
similar to that of the familiar SM, the {\em calculable} one-loop contributions
to gauge couplings giving unification to within $\sim 20 \%$. 
But unlike the SM, 
there is a natural source of small, but volume-enhanced 
bulk
threshold corrections needed to fit the observed couplings. 
(iii) The fact that the
unification scale is not far from the Planck scale is not an accident, 
but follows from
the RS1 approach to solving the hierarchy problem. (iv) There is a natural
solution to the problem of baryon-number violation which puts our models on a very
similar footing to the SM.
(v) It is a remarkable feature of
the RS1 mechanism that it has a purely $4D$ dual description, given by the AdS/CFT
correspondence \cite{adscft, rs2cft, nima1, rs1cft},  
in which the Higgs is a TeV scale composite of a
strongly coupled large-$N$ conformal field theory. All of the above
features of unification have simple $4D$ interpretations which clarify their more
general applicability for theories of strongly 
interacting Higgs sectors. Yukawa coupling
unification is not studied in this paper.

Beyond its intrinsic interest,
we believe that the study of non-supersymmetric unification in the RS1 scenario is
of value in coming to a more balanced view of weak scale SUSY.
The attractions of supersymmetric grand unified theories (GUTs)
and the absence so far
of definitive
experimental confirmation poses tough questions for particle physicists.
How strong a hint is the
seamless fit of unification into weak scale SUSY that these ideas are actually
realized in Nature? 
Now {\em qualitatively}, RS unification shares many of 
the features of supersymmetric
unification. 
{\em Quantitatively}, RS unification is attractive but 
does not match the precision
of supersymmetric unification, where the {\em calculable} contribution to 
running of gauge couplings gives unification to within $\sim 4 \%$ compared to
$\sim 20 \%$ for the SM and our RS1 scenario. 
In both supersymmetric GUTs and RS GUT, the small discrepancies 
in unification can {\em naturally} be attributed to UV-{\em sensitive}, but
parametrically small, threshold corrections.
We see that the phenomenological support for 
unification is not an all or nothing pointer towards SUSY. Most of this support
is present in the SM, and can be maintained by sensible
non-supersymmetric approaches to the hierarchy problem like RS1.
It is only the {\it extra} precision in the fit of  supersymmetric
unification that has to be  
an accident in order for weak scale SUSY to be false, but 
much less of an accident than if only SUSY could combine unification with a
solution to the hierarchy problem.

There are important distinctions
between unification in RS1 and unification in unwarped extra dimensions
\cite{antoniadis, ddg, nima2, orbifoldnew, bhn, jmr}. 
Reference \cite{antoniadis}
proposed high-scale 
unification with gauge fields propagating in TeV scale extra dimensions, with 
some of the gauge symmetry and $N = 4$ SUSY broken by orbifold boundary 
conditions. The bulk $N = 4$ SUSY protects logarithmic running above the 
compactification scale. In both this scenario and in the RS model the 
non-renormalizable higher-dimensional effective field theory must be 
UV-completed not far above the TeV scale. In spite of that, in the RS scenario 
we consider, we are able to controllably study unification at much higher scales 
within effective field theory. 
Reference \cite{ddg} found {\em low}-scale unification by studying
supersymmetric extra dimensional models . Gauge coupling unification was
{\em not} correlated in a simple way with the observed strength of gravity because
of the absence of logarithmic running. Reference \cite{nima2} studied 
a toy model of unification in the context of large 
(gravity-only) extra dimensions. 
Within specific superstring theory models, unification was achieved by
classical logarithmic potential effects which closely resemble effects of $4D$
logarithmic running, in particular, gauge coupling unification {\em was}
correlated
with the observed strength of gravity as we will also find. However, whereas in
reference \cite{nima2}, the fact of unification depends entirely on stringy 
UV details, in the unification scheme we study, the dominant ingredients are 
calculable in the
effective field theory.
References \cite{orbifoldnew, bhn, jmr} studied {\em high}-scale unification, but with 
compactification scales close to the unification scale. In our models, because
of warping, the Kaluza-Klein (KK) mass scale is far lower than 
the unification scale.
However, our mechanism for avoiding excessive proton decay by 
breaking the GUT symmetry
with boundary conditions is similar to mechanisms of references 
\cite{ddg, orbifoldnew, bhn}.

Several issues of the RS
set-up relevant for our paper
already appear in the literature.
Unification in an RS setting was first 
studied in a {\em supersymmetric} 
version in reference \cite{pomarolprl} (and recently in reference \cite{gns}), 
SUSY allowing for a different placement of matter and Higgs
fields from ours. 
Reference 
\cite{pomarolprl} pointed out the basic feature of logarithmic ``running'' 
in RS1 
{\it above} the lightest KK mass,  on which our work relies. 
Reference \cite{lisa} first proposed
{\em non}-supersymmetric 
coupling unification in the RS solution to the hierarchy problem and
studied the requisite logarithmic running. 
It differs from our work in that it focused more on the case where there is 
no explicit
unified gauge group, in
the proposal for baryon number conservation and in the
approximations made. 
Reference \cite{contino} also studied
logarithmic running and non-supersymmetric GUTs (including the CFT
dual description), but did not discuss quantitative
unification 
and the proton decay problem.
References \cite{choi1, gr, us1, deconstruct, choi2, us2} 
further studied gauge coupling renormalization in RS1, relevant
for unification, from various points
of view. In particular, we laid the foundations for our approach to 
differential running of gauge couplings in reference \cite{us1}.

While having SM gauge fields propagate in the RS bulk is good for
unification, there have been studies pointing out several unattractive
phenomenological features \cite{hewettpomarol, pheno}. 
In particular these studies suggest that unless the warped down
fundamental scale is of order $10$ TeV or higher, KK gauge bosons would
make excessive contributions to electroweak precision and
compositeness observables. 
Such a high fundamental scale would render the Higgs and the weak scale somewhat
fine-tuned. Actually, we find the phenomenology of the RS1 scenario with 
bulk gauge fields
very intriguing and subtle and believe that the severe 
constraints found thus far can be
considerably weakened by simple
model-building. Interesting as it is
from both theoretical 
and  
experimental 
points of view, 
we will not delve into these topics of weak scale 
constraints and signals
here, but will reserve that for a later
paper \cite{phenous}. 
For present purposes the reader can take the warped down 
fundamental scale to be a few TeV, 
the exact number not being important given the precision in
coupling unification we will attain.

The outline of our paper is as follows. In Section \ref{review} 
we review the standard
unification in the minimal SM and in the
minimal supersymmetric standard model (MSSM), 
so as to lay the foundation for our study of 
RS unification. In Section \ref{module}, 
we present a minimal module for RS unification, discuss its
features and compare
it with the more familiar unification scenarios. In Section \ref{matter}, 
we discuss SM
fermions living in the extra dimensional bulk and give a mechanism that leads to
accidental (but, anomalous) 
baryon-number conservation, very similar to that of the ordinary
SM. Thus, our basic module can be completed into realistic models.
In Section \ref{cft}, we give dual 
CFT interpretations 
of our models and their features.
In section \ref{conclude}, we provide our conclusions.

\section{Review of $4D$ unification}
\label{review}

\subsection{Gauge couplings in a general GUT}

The structure of a $4D$ GUT at one-loop is
\begin{eqnarray}
\alpha ^{-1} _i \left( M_Z \right) & = & \alpha ^{-1} _{GUT} + \frac{b_i}
{2 \pi} \log \frac{ M_{GUT} }{M_Z} + \Delta _i,  
\label{4dgut}
\end{eqnarray}
where 
$M_{GUT}$ is the scale at which the GUT symmetry is broken and
$\alpha_{GUT}$ is the unified gauge coupling at that scale.
The running of the gauge couplings
in the effective theory below $M_{GUT}$ is given by the $b_i$ terms, whereas  
the 
$\Delta _i$'s are threshold corrections from 
the GUT and weak scales and the effects of Planck-suppressed
higher-dimensional operators.\footnote{For simplicity,
we assume a ``desert'', 
i.e., no other mass thresholds between $M_{GUT}$ and the weak scale.}

Assuming unification, we can
eliminate $\alpha _{GUT}$ and $M_{GUT}$ from the above three
equations 
to get 
%
\begin{eqnarray}
\left( \alpha ^{-1} _1 \left( M_Z \right) - \Delta _1 \right) 
\left( b_2 - b_3 \right) + 
\left( \alpha ^{-1} _2 \left( M_Z \right) - \Delta _2 \right) 
\left( b_3 - b_1 \right)
+  \left( \alpha ^{-1} _3 \left( M_Z \right) - \Delta _3 \right) 
\left( b_1 - b_2 \right) & = & 0 \nonumber \\
\end{eqnarray}
and hence we 
can solve for the following combination of $\Delta _i$'s in terms of
the observed gauge couplings and $b_i$'s: 
\begin{eqnarray}
\Delta & \equiv &
\Delta _1 \frac{ b_2  - b_3 }{ b_1 - b_2 } 
+ \Delta _2 \frac{ b_3 - b_1 }{ b_1 - b_2 } + \Delta _3 \nonumber \\
 & = & \alpha _1 ^{-1} \left( M_Z \right) \frac{ b_2  - b_3 }{ b_1 - b_2 } + 
\alpha _2 ^{-1} \left( M_Z \right) \frac{ b_3 - b_1 }{ b_1 - b_2 } + 
\alpha ^{-1} _3 \left( M_Z \right). 
\label{Delta}
\end{eqnarray}
We will use $\alpha _1 \left( M_Z \right) \approx 0.017$, 
$\alpha _2 \left( M_Z \right) \approx 0.034$ and
$\alpha _3 \left( M_Z \right) \approx 0.117$ \cite{pdg}.

\subsection{SM}
\label{smgut}
\begin{figure}[t]
\centering
\epsfig{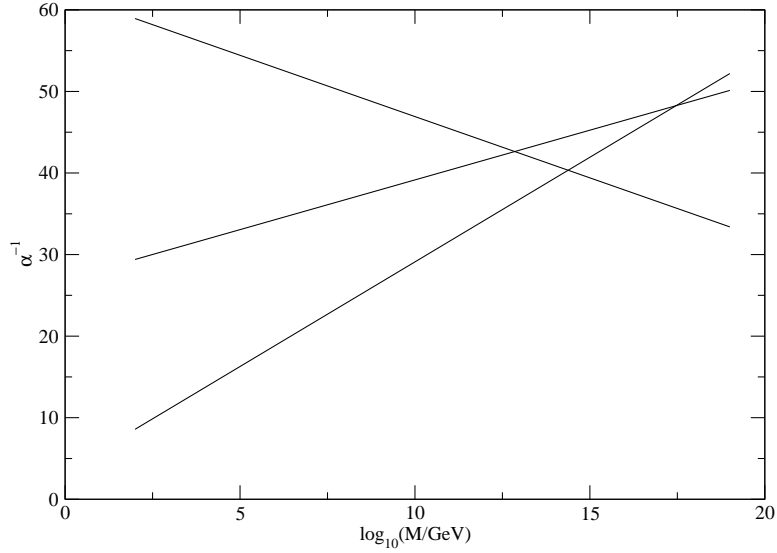}
\caption{Running couplings ($\alpha^{-1}$) with 
respect to energy scale ($M$) for the SM.}
\label{sm}
\end{figure}
We plot the running of gauge couplings in Fig.~\ref{sm} for $\Delta _i = 0$.
We see that the three gauge couplings do not quite meet, although 
they come close
to each other at energies of $O \left( 10^{15} \right)$ GeV.
Using $b_i^{SM} = (41/10, -19/6, -7)$ (one-loop
contribution of gauge bosons + matter + 1 Higgs doublet)
in Eq.~(\ref{Delta}), we get $\Delta_{SM} \approx -5$ if we impose 
unification.\footnote{We get also a range of $M_{GUT} \sim 10^{13}-10^{18}$ GeV
depending on the values of individual $\Delta_i$'s.}

We can compare the required (threshold) correction
to the typical size of
{\em differential} running contribution which is dominated by gauge boson loops. 
Their contributions
are given by $11/ (6 \pi) \log \left( M_{GUT} / M_Z \right)
\times O(1)$ group theory factors. 
We see that requiring $\Delta _{SM} = -5$
corresponds to $\sim 20 \%$ correction to differential running.
Let us see if such $\Delta$ is reasonable. There are
two types of threshold corrections:
loops of particles at $M_{GUT}$ 
which contribute $\sim 1 / \left( 2 \pi \right) \times$
number of particles at $M_{GUT}$
(no large logarithms) 
and Planck-suppressed
operators 
$\sim \Sigma F^2 / M_{Pl}$ (where
$\langle \Sigma \rangle$ breaks the GUT)
which contribute $\sim 4 \pi \times M_{GUT} / M_{Pl}
\sim 10^{-1}$.
Assuming no unnaturally large parameters (for example, the number of particles
at $M_{GUT}$ is not too large),
both corrections are much smaller than the required $\Delta$.
Of course, other than this significant problem,
the SM does not protect the weak scale from the unification scale. We will see 
that in
RS unification, there is a natural source for the requisite $\Delta$ and 
protection for large hierarchies.

\subsection{MSSM}
%
%
As is well-known, the fit to perturbative unification in MSSM is very good at 
one-loop 
and in fact, a {\em two}-loop analysis is warranted. 
The central ingredient in the success of 
supersymmetric GUTs is the addition of the extra Higgs(inos) and gauginos.
%
%
The extra {\em differential} running due to these particles provides the 
$\sim 20 \%$ contribution (on top of the SM differential running)
required to achieve unification.
%
%
When the analogous two-loop treatment of 
Eq.~(\ref{Delta}) is done, one finds $\Delta \sim 1$ (see, for example, 
reference \cite{raby}). 
However, unlike the SM, this 
value of $\Delta$ is naturally consistent with the
expected size of GUT and
weak scale
threshold
corrections. 

\section{A minimal module for RS1 unification}
\label{module}

We first consider a minimal module for RS unification which consists of 
$SU(5)$ unified gauge
fields in the bulk, along with scalar fields required for
Higgsing the GUT down to the SM gauge group.
We will include 
other
matter 
in the next section.
The bulk action is
\begin{eqnarray}
S_{bulk} & = & \int d^4 x dy \sqrt{-G} \; \hbox{Tr} 
\left( - \frac{1}{4 g_5^2} F_{MN}
F^{MN} + | D_M \Sigma |^2 + V_{bulk} (\Sigma) + \frac{a_{bulk}}{ \sqrt{\Lambda} }
\Sigma F_{MN} F^{MN} \right). \nonumber \\ 
\end{eqnarray}
Here, $F_{MN}$ is the $SU(5)$
gauge field strength, $\Sigma$ is a scalar field transforming under $SU(5)$
and $\Lambda$ is a Planckian scale where the RS effective field theory
becomes strongly coupled.
We will take $A_{\mu}$ and $\Sigma$ to be even and $A_5$ to be odd
under the orbifold symmetry.
The potential $V_{bulk} (\Sigma)$ is such that
$\Sigma$ acquires a vev, $\langle \Sigma 
\rangle \equiv v^{3/2}$ 
which breaks $SU(5)$ gauge symmetry to the SM gauge symmetry.
In addition, we have the brane-localized action:
\begin{eqnarray}
S_{UV (IR)} & = & \int d^4 x \sqrt{ -g_{UV(IR)}  }
\; \hbox{Tr} \left (- \frac{1}{4} \tau_{UV(IR)}
F_{\mu \nu} F^{\mu \nu} + \sigma_{UV(IR)} | D_{\mu} \Sigma |^2 \right. 
\nonumber \\
 & & \left. + V_{ UV (IR) } ( \Sigma ) + \frac{a_{UV (IR) }}{ \sqrt{\Lambda ^3} }
\Sigma F_{\mu \nu} F^{\mu \nu} \right ). 
\end{eqnarray}
We assume that brane-localized terms 
are of order loop processes involving bulk couplings as well as
Planck-suppressed effects. 
They are therefore neglected within one-loop Feynman diagrams.

At one-loop order, the vacuum polarization diagrams are UV divergent and require
renormalization of $1/g_5^2$ and $\tau_{UV (IR)}$. This was first
discussed in references \cite{gr, us1}. Henceforth, we will work with the renormalized
couplings denoted by an extra subscript $R$.

Due to the
Higgs mechanism, $12$ (real) degrees of freedom of $\Sigma$ are eaten 
by the $X$, $Y$ gauge bosons which get a mass of $M_{GUT}^2 \sim g_5^2 v^3$. 
We will take $M_{GUT} / k, M_{GUT} / \Lambda \ll 1$.
The
remaining components of $\Sigma$ are physical -- we will assume that they
have a mass, $M_{\Sigma}$ of $O \left( M_{GUT} \right)$. We then calculate
the $4D$ low energy gauge couplings at one-loop order
by a similar reasoning
to our earlier analysis of scalar (massless and massive) QED \cite{us1}
(also see references \cite{pomarolprl, lisa, choi1, gr, contino, deconstruct, choi2}). 
The massless particles in the
loop are the SM gauge bosons, whereas the massive particles are the $X$, $Y$ 
gauge bosons {\em and} the physical components of $\Sigma$.
We get 
%
\begin{eqnarray}
\alpha _i ^{-1} (M_Z) & =  & \frac{\pi r_c}{\alpha _{5 \; R} (k)} + 
O \left( \frac{1}{2 \pi} \right)
\nonumber \\
 & & - \frac{55}{6 \pi} \left( \log \frac{k}{ M_{GUT} } 
+ \xi ^{\prime} k \pi r_c \right)
+ \frac{ C ( {\Sigma} ) }{12 \pi} \left( \xi k \pi r_c + \log \frac{k}{ M_{GUT} }
\right) 
\nonumber \\
 & & + \frac{ b_i^{RS} }{2 \pi} \;
\log \frac{ M_{GUT} }{M_Z}
+ O \left( \frac{1}{2 \pi} \times
\frac{ M^2_{GUT} }{ k^2 } \right) _i \; k \pi r_c 
\nonumber \\
 & & +
a_{i \; bulk} k \pi r_c 
\frac{4 \pi}{ g_5 \sqrt{k} } \frac{ M_{GUT} }{ \sqrt{ k \Lambda} }. 
\label{rsgut1}
\end{eqnarray}
A brief explanation is as follows (for details, see \cite{us1}).
The 
first 
term on the RHS of Eq.~(\ref{rsgut1})
(with $\alpha_5 = g_5^2 / \left( 4 \pi \right)$)
is  {\em universal} for $i = 1, 2, 3$. 
The second line is a {\em universal} (for $i = 1, 2, 3$) one-loop contribution 
-- the first term is from gauge bosons and the second term is from
the $\Sigma$ field (assuming that the mass of {\em all} components
of $\Sigma$ is $\sim M_{GUT}$), where $C ({\Sigma})$ is the 
Dynkin index
of the representation of $\Sigma$ and $\xi$, $\xi ^{\prime}$ are $O(1)$ constants.
The third line is the remaining, {\em non}-universal part of the one-loop contribution 
-- the first term is from gauge bosons,
where 
\begin{eqnarray}
b_1^{RS} & = & 0, \nonumber \\ 
b_2^{RS} & = & - \frac{22}{3}, \nonumber \\ 
b_3^{RS} & = & - 11
\end{eqnarray}
and 
the $O \big[ 1 / \left( 2 \pi \right) \times
\left( M^2_{GUT} / k^2 \right) \big]_i \; k \pi r_c$
terms are one-loop threshold contributions arising from GUT scale splittings in 
$5D$ masses. We take these splittings to be $O \left( M_{GUT} \right)$. These
contributions can be precisely computed in any particular $5D$ GUT model, but we will 
only need their basic size here, assuming $M_{\Sigma} \sim O \left( M_{GUT} \right)$.
The $O \big[ 1 / \left( 2 \pi \right) \big]$ terms 
(on the first line) are effects which are of one-loop
order and not enhanced by $r_c$ or $\log \left( M_{GUT} / M_Z \right)$ 
or $\log \left( k / M_{GUT} \right)$.
The $a_i$-terms in the last line 
are the contributions from the higher-dimensional operator:
$a_{i \; bulk} = \gamma_i a_{bulk}$, where 
$\gamma_i$'s ($\sim O(1)$) depend on the representation of $\Sigma$.
Factors of $k$ have been inserted for later convenience.

\begin{figure}[t]
\centering
\epsfig{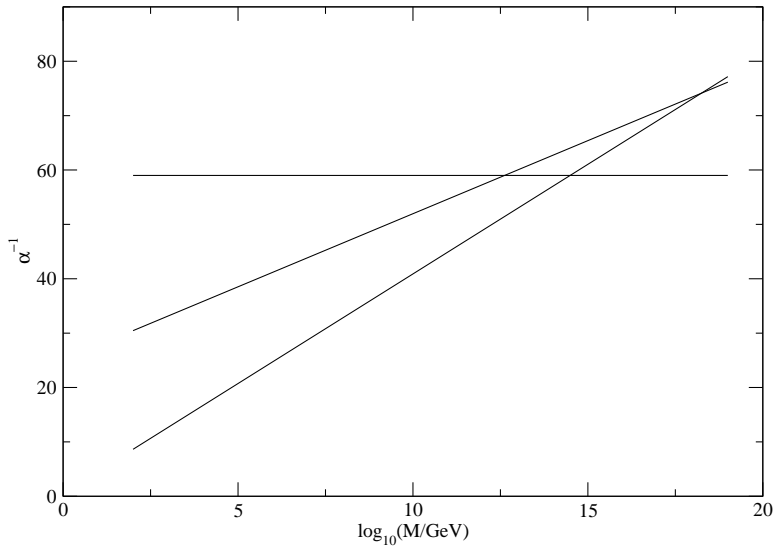}
\caption{Running couplings ($\alpha^{-1}$) 
with respect to energy scale ($M$) for RS1 with
$\Delta_i=0$.}
\label{rswithoutdelta}
\end{figure}

We now come to the central point of our paper by relating RS unification to the
structure
of $4D$ GUTs as given by Eq.~(\ref{4dgut}). The significance of separating
non-universal pieces in the RHS of Eq.~(\ref{4dgut}) 
is that the $b_i$ are calculable 
from low-energy degrees of freedom, whereas $\Delta_i$ depend on details of
UV physics at GUT and Planckian scales. Since these details are unknown
{\it a priori}, Eq.~(\ref{4dgut}) is only useful if the 
$\Delta_i$ are naturally small. Remarkably, 
neglecting unenhanced $O \big[ 1 / \left( 2 \pi \right) \big]$ terms,
Eq.~(\ref{rsgut1}) can be put
in just this form:
\begin{eqnarray}
\alpha ^{-1}_{GUT} 
& 
= 
& 
\frac{\pi r_c}{\alpha _{5 \; R} (k)} 
\nonumber \\
 & & 
- \frac{55}{6 \pi} \left( \log \frac{k}{ M_{GUT} } 
+ \xi ^{\prime} k \pi r_c \right)
+ \frac{ C ( {\Sigma} ) }{12 \pi} \left( \xi k \pi r_c + 
\log \frac{k}{ M_{GUT} }
\right)
+ O \left( \frac{1}{2 \pi} \times
\frac{ M^2_{GUT} }{ k^2 } \right) k \pi r_c 
\nonumber \\ 
b_i  
& 
= 
& 
b_i^{RS} \nonumber \\
\Delta _i 
&
=
& 
a_{i \; bulk} k \pi r_c 
\frac{4 \pi}{ g_5 \sqrt{k} } \frac{ M_{GUT} }{ \sqrt{ k \Lambda} }
+
O \left( \frac{1}{2 \pi} \times
\frac{ M^2_{GUT} }{ k^2 } \right) _ i \; k \pi r_c. 
\label{rsgut2}
\end{eqnarray}
Note that $\Delta_i$ are naturally suppressed if the effective field theory is 
sufficiently weakly coupled at the GUT scale, i.e., 
$M_{GUT} / k, M_{GUT} / \Lambda \ll 1$ as we are considering.
These are the same qualitative reasons that suppress  $\Delta_i$ in the SM
and MSSM. Also,
note that the contribution with 
$b_i^{RS}$'s (which is roughly the running due to zero-modes of gauge fields) is 
the same as in the SM as far as {\em differential} running goes except 
for the very small contribution of
Higgs scalar (running due to fermions in the
SM is {\em universal}).
In the limit $\Delta_i \rightarrow 0$, 
a plot of Eq.~(\ref{4dgut}) with $b_i^{RS}$ is
shown in Fig.~\ref{rswithoutdelta}. Note that 
this is very close to the SM as far as differential
running goes (cf. Fig.~\ref{sm}). 
Fig.~\ref{rswithoutdelta} strongly suggests high-scale
unification expected within the RS 
philosophy of no very large fundamental hierarchies.

Although $\Delta_i$ are suppressed for small $M_{GUT}$, they are  
enhanced in RS1 by the extra-dimensional volume (i.e.,
$k \pi r_c$) compared to the SM.
This is the key distinction with $4D$ SM GUTs.
While the $\Delta_i$ of SM GUTs could not {\em naturally} account for the
even mild discrepancy in unification of Fig.~\ref{sm}, 
the volume-enhanced $\Delta_i$
of Eq.~(\ref{rsgut2}) can easily do so in 
Fig.~\ref{rswithoutdelta} as we now show.

\begin{figure}[t]
\centering
\epsfig{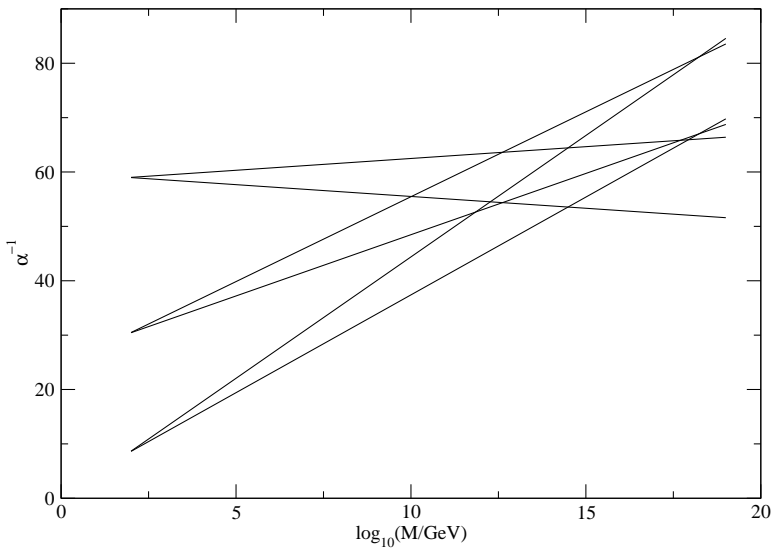}
\caption{Running couplings ($\alpha^{-1}$) 
with respect to energy scale ($M$) 
for RS1 with $\Delta_i=\pm O(10 \%)$ of differential
running contributions in the SM.}
\label{rswithdelta}
\end{figure}

The crucial observation is that
$k \pi r_c \sim
\log \big[ O \left( M_{Pl} \right) \big] / \hbox{TeV}$,
required in RS1 to solve the Planck-weak hierarchy problem.
According to the RS philosophy of no very large hierarchies, we have
$k \pi r_c, \log \left( k / M_Z \right) \sim \log \left( M_{GUT} / M_Z \right) 
+ O(1)$.  
Thus,
keeping only the single large logarithm and dropping non-log-enhanced terms, 
we get
\begin{eqnarray}
\Delta _i & = & \log \frac{ M_{GUT} }{M_Z} 
\Big[ a_{i \; bulk} \frac{4 \pi}{ g_5 \sqrt{k} }
\frac{ M_{GUT} }{ \sqrt{k \Lambda} } +
O \left( \frac{1}{2 \pi} \times \frac{ M^2_{GUT} }{k^2} \right) _i \Big] 
\nonumber \\
\alpha ^{-1}_{GUT} & = & \log \frac{ M_{GUT} }{M_Z} 
\Big[ \frac{1}{ k \; \alpha_{5 \; R} (k) } +
\frac{-55}{6 \pi} 
\xi^{\prime} 
+ \frac{ C ( {\Sigma} ) }{2 \pi} 
\xi
+ O \left( \frac{1}{2 \pi} \times \frac{ M^2_{GUT} }{k^2} \right)
\Big].
\label{rsgut3}
\end{eqnarray}
From Eqs.~(\ref{4dgut}) and (\ref{rsgut3}),
we see that,
due to the log-enhancement, $\Delta _i$ in
RS1 {\em resemble} a change in $\beta$-function 
coefficient (i.e., change in {\em running} of the gauge coupling) 
\begin{equation}
\Delta _i = \log \frac{ M_{GUT} }{M_Z} \times  \left(
 \frac{ \delta b_i }{2 \pi} \right), 
\end{equation}
even though
their origins are bulk {\em threshold} 
corrections and Planck-suppressed operators. 
We see that, with {\em modestly} small parameters, 
$M_{GUT} / k, M_{GUT} / \Lambda$, 
the
RS1 $\Delta$ 
can be $\sim 20 \%$ compared to the SM {\em differential} running --
as mentioned earlier,
the right size of correction to the SM in order to achieve unification. 
Note that if the data had supported much lower unification scale with a 
similar requirement for $\Delta$, then 
our RS unification would have been falsified since
$\Delta$ would have been extremely small (see Eq.~(\ref{rsgut3})).
We illustrate this phenomenon in Fig.~\ref{rswithdelta}, including the effect of 
$\Delta_i$.
We allow
$\Delta_i$ to vary between $\pm 10 \%$ of the typical
differential running contribution in the SM 
$\sim 11 / \left( 6 \pi \right) \log \left(
M_{GUT} / M_Z \right)$ (cf. section \ref{smgut}). 
Note that all three ``cones'' in Fig.~\ref{rswithdelta} 
do overlap consistent with
unification. We stress that RS GUT theory does {\em not} predict that
$\Delta _i$ corresponds to a $10 \%$ effect compared
to the calculable running. Rather, the data show in Fig.~\ref{rswithdelta} 
that $O(10) \%$ effect
is sufficient for unification. 
What we have shown is that such (small) effects are 
{\em naturally} present in RS GUTs. Therefore, the near unification
of couplings seen in Fig.~\ref{rswithoutdelta} 
is not just an accident. 

Note that in {\em flat} extra dimensions
with SM gauge fields, the effect of the higher-dimensional operator
is similarly volume-enhanced, but in that case
(unlike RS1),
$r_c$
is not related to the Planck-weak hierarchy.

It is interesting to compare this RS unification
to the unification in the MSSM. 
As mentioned above,
MSSM also has modest ($\sim 20 \%$), 
but important corrections to 
({\em differences} of) $\beta$-function coefficients due 
to additional Higgs(inos)and
gauginos.
The crucial difference between MSSM and RS1 is that
in the MSSM, the modifications {\em based on minimality} are completely fixed,
whereas in RS1 (at this stage of its theory), 
while threshold corrections are parametrically small, they 
are determined by {\em independent} parameters of the effective field theory.

As a concrete illustration of unification in 
the intersection of the three cones in Fig.\ref{rswithdelta}, 
we consider $\Sigma$ to transform as
a ${\bf 75}$ under $SU(5)$. We have checked
that 
we can get very good 
unification of gauge couplings in this case
for $M_{GUT}, M_{\Sigma} \sim 10^{16}$GeV,
$k \sim 10^{17}$GeV, $\Lambda \sim 10^{18}$GeV and $a_{bulk}  \sim 1$.

\section{Bulk matter and proton decay}
\label{matter}

In any theory with unification there is
a potential problem with proton decay mediated by $X$, $Y$ gauge bosons. 
In the theory we are dealing with the problem is even worse because 
the lightest such states in the spectrum have masses $\sim$ TeV.
Also, there can be higher-dimensional 
operators which violate baryon-number which near the 
IR brane are suppressed
by only the TeV scale.
In this section, we will show how both these problems can be naturally solved,
allowing us to complete our basic module into realistic models.

Before addressing these issues, we must decide where matter fields are located
in the extra dimension. In order to solve the hierarchy problem, we will keep 
the Higgs boson confined to the TeV brane. Reference \cite{lisa} studied the issue
of baryon-number conservation when fermions are also confined to the TeV brane. 
However, this situation 
has some phenomenological
problems forcing us to push up the fundamental scale
\cite{pheno}.
Instead, we will consider quarks and leptons as free to propagate 
in the bulk. 
In a future paper \cite{phenous},  
we will show how this placement can significantly
weaken the phenomenological constraints. 
Proton decay in {\em supersymmetric} RS1 scenario was discussed in 
references \cite{pomarolprl, gns}, but with a different placement of the Higgs.

$5D$ theories are non chiral, chirality being
achieved by means of the orbifold. 
This scenario allows for odd bulk mass terms for fermions of the form 
$c k \varepsilon(y)\bar{\Psi}\Psi$ (where $\varepsilon$ is the sign function) 
which is compatible
with having a massless chiral mode upon compactification. Essentially
$c$ controls the localization of the wavefunction of the massless mode.
In the warped scenario, for
$c>1/2$ ($c<1/2$) the zero mode is localized near the Planck (TeV) brane,
whereas for $c = 1/2$, the wave function is {\em flat}. 
When matter fermions are in the bulk,
there is an elegant way of avoiding baryon-number
violation mediated by $X$, $Y$ states   
using boundary conditions to break $SU(5)$ 
\cite{bhn}
in addition to
our bulk breaking 
by $\Sigma$.\footnote{Gauge symmetry breaking by boundary conditions
was introduced in string phenomenology in reference \cite{orbifoldold}.}
This involves replacing the usual $\mathbb{Z}_2$ orbifold by a 
$\mathbb{Z}_2\times\mathbb{Z}'_2$ 
orbifold, where $\mathbb{Z}_2$ corresponds to reflection about the Planck brane
and $\mathbb{Z}'_2$ corresponds to reflection about the TeV brane.
Under this orbifold, the SM gauge fields are $(+,+)$, 
whereas $X_{\mu}$ and $Y_{\mu}$ can either
be $(+,-)$ 
(breaking $SU(5)$ on the TeV brane) or $(-,+)$ (breaking $SU(5)$
on the Planck brane). 
When putting fermions in the bulk, 
one is forced to obtain quarks and lepton zero-modes
from different $SU(5)$ bulk multiplets \cite{bhn, jmr}. In particular, $X$, $Y$
exchanges cannot connect SM quark zero-modes to SM lepton zero-modes
\cite{bhn}.
For the case 
of the fermions only the left handed chirality will be discussed since the
right chirality is projected out. 
We have two (left-handed) $\mathbf{\bar{5}}$'s and two (left-handed) 
$\mathbf{10}$'s per generation in such a way that: 
\begin{eqnarray}
& \mathbf{\bar{5}}_1=\mathbf{L}_1+d^{c}_1 
\quad \mathbf{\bar{5}}_2=L_2+\mathbf{d}^{c}_2& \nonumber \\
&\mathbf{10}_1=\mathbf{Q}_1+e^{c}_1+u^{c}_1 \quad 
\mathbf{10}_2=Q_2+\mathbf{e}^{c}_2+\mathbf{u}^{c}_2,&
\end{eqnarray}
\noindent
where the particles in boldface are the ones to have zero modes, i.e. 
to be $(+,+)$. The extra 
fields needed to complete all representations
can either be $(+,-)$ 
(if breaking of $SU(5)$ is on the TeV brane) or
$(-,+)$ (if breaking is on the Planck brane), i.e.,
depending on the choice we make for $X$ and $Y$ bosons, so these 
extra fermionic fields will not have massless
modes upon compactification. 

An important consideration in unification is the effect of fermion loops.
For the 
case of $(+,-)$ 
($SU(5)$ breaking on the TeV brane), these loops cannot contribute to the large
logarithm appearing in the {\em differential} running. But, for the 
$(-,+)$ case ($SU(5)$ breaking on the 
Planck brane), this is not automatically true (using the
results of \cite{choi2}). In this case, fermions will only
give universal contribution at one-loop if $c \stackrel{>}{\sim} 1/2$ or
if $c_{5_1} \approx c_{5_2}$ and
$c_{10_1} \approx c_{10_2}$. We will study the phenomenology of both cases in 
a future paper \cite{phenous}. 

In order to control proton decay from higher-dimensional operators 
suppressed by the TeV scale, one can impose an extra symmetry $\tilde{U}(1)$ 
\cite{gns} under which
each field has the following charges\footnote{Usual
baryon number can{\em not} be used since it does not commute with $SU(5)$.}:
\begin{equation}
\mathbf{\bar{5}}_1(-1)\quad\mathbf{\bar{5}}_2(4)\quad\mathbf{10}_1(-3)
\quad\mathbf{10}_2(2)\quad H(-1).
\end{equation}
\noindent
Here, the Higgs doublet, $H$
has $Y = -1/2$. Since $Q_{\tilde{U} (1)} = 2 Y - 10 B$
(where proton has $B = 1$), after electroweak 
symmetry breaking, $U(1)_B$ symmetry remains to protect baryon number. 

The $\tilde{U} (1)$ symmetry has $\big[ \tilde{U} (1) \big] ^3$, 
SM$\times \big[ \tilde{U} (1) \big]^2$ and (SM)$^2 
\times \tilde{U} (1)$ anomalies after compactification. 
One possibility is to keep such a symmetry global (and anomalous) \cite{gns},
but there is a danger that quantum gravity 
effects do not respect such a symmetry. 
Instead, here, we will {\em gauge} the $\tilde{U} (1)$ symmetry in the bulk
($U(1)_B$ bulk gauge fields in the context of large extra dimensions
were introduced in references \cite{nima3}), adding
brane-localized ``spectators'' to
cancel the $\tilde{U} (1)$ anomalies. Indeed, only this gauged 
scenario has a {\em simple} dual CFT interpretation as
we discuss later. 
Under $\mathbb{Z}_2\times\mathbb{Z}'_2$, we will take the 
$\tilde{U} (1)_{\mu}$ gauge boson to be $(+, +)$.
We will 
break this gauge symmetry at the Planck scale by giving the gauge field
a Planckian mass localized on the Planck brane (for example,
due to the condensation of a SM singlet ``baryonic'' scalar). Thus, any
higher-dimensional
baryon-number violating operators will be
have to be localized on the Planck brane and hence 
will be suppressed by the Planck scale.
So, we have accidental baryon-number conservation as in the SM.

A possible choice of colorless spectators
are two doublets of $SU(2)$ with $Y=0$ and $Q_{\tilde{U}(1)}=15$, two 
$SU(2)$ singlets with
$Y=1/2$ and $Q_{\tilde{U}(1)}=-14$ and two $SU(2)$ singlets with
$Y=-1/2$ and $Q_{\tilde{U}(1)}=-16$.\footnote{A similar set
of spectators was used in 
reference \cite{cm} to cancel anomalies involving baryon-number
which is equivalent to canceling $\tilde{U} (1)$ anomalies.} 
The spectators are {\em not}
in complete $SU(5)$ multiplets
and so will have to be localized on the $SU(5)$-breaking brane.
If we break $SU(5)$ by orbifolding on the Planck brane then these spectators
will be localized there and can get a 
Planckian mass after $\tilde{U}(1)$ is broken
(by coupling to the scalar vev which breaks $\tilde{U}(1)$). 
Without the $\tilde{U} (1)$ gauge field zero-mode and 
these spectators, the low-energy
theory is not protected by an 
exact baryon {\em gauge} 
symmetry, but has accidental {\em 
anomalous} global baryon-number (just like the SM) below
the Planck scale.
If $SU(5)$ is broken on the TeV brane, then the spectators have to be 
localized there too and
their mass will be of electroweak origin by Yukawa
couplings to the Higgs. 
Phenomenologically, as long as masses of these spectators
are over $100$ GeV, there is no problem from present bounds on 
direct detection \cite{pdg}. Moreover, the contributions to
electroweak observables, such as the $S$-parameter, 
are within errors (see, for example, the review \cite{S}). 
Again, without $\tilde{U} (1)$ gauge field zero-mode, baryon-number
is not an exact symmetry, but is an 
accidental symmetry ({\em non}-anomalous) below the
Planck scale.

Finally, we discuss the Higgs doublet-triplet splitting problem. To solve the
hierarchy problem using the warp factor, the Higgs doublet will have to
be localized on the TeV brane.
If we break $SU(5)$ on the TeV brane, then we need {\em not} add a color 
triplet partner
for the Higgs doublet. 
In the case of $SU(5)$ breaking on the Planck brane, there is a 
Higgs triplet accompanying the Higgs doublet on the TeV brane.  
Let us see if this causes phenomenological problems, for example,
for
proton decay.\footnote{Since the Higgs doublet and 
triplet are localized on the TeV
brane, they will contribute to any running of gauge couplings 
only {\em below} the
TeV scale. Therefore, large logarithmic running is not affected.} 
We need a TeV brane-localized Yukawa coupling $L_1 e_2^c H_2$
to give the electron a mass,
where $H_2$ is the Higgs doublet. By $SU(5)$ gauge invariance, there
is a coupling $d^c_1 u^c_2 H_3$, where $H_3$ is the Higgs triplet.
However, $d^c_1$ does not have a zero-mode (although $u^c_2$ does)
so that $H_3$ does {\em not}
couple SM fermions to each other (just 
as in the case of $X$, $Y$ gauge bosons) and hence
the presence of a light Higgs triplet does {\em not} lead to proton decay.

Another issue is how to ensure that $H_3$ has a {\em positive}
(mass)$^2$ especially since $H_2$ should have a {\em negative} (mass)$^2$
to break electroweak symmetry
and $SU(5)$ symmetry relates the two masses. 
We can use couplings to the $\Sigma$ vev
(which breaks $SU(5)$ symmetry) for this purpose -- in the absence of 
the $\Sigma$ vev,
the two masses will have to be equal at
{\em tree}-level by $SU(5)$ symmetry since $SU(5)$ is broken
{\em only} on the Planck brane in that case.
Also, 
$4D$ {\em radiative} corrections can split the two masses.
 

\section{CFT interpretation}
\label{cft}

\subsection{The minimal module}
\label{cftminimal}

We begin with a dual $4D$ CFT interpretation of 
the minimal module of RS GUTs, i.e.,
with no SM matter or Higgs doublet and no orbifold breaking of $SU(5)$.
Earlier discussions of RS bulk gauge fields from the dual point of view
appear in references \cite{nima1, gr, us1, contino, us2}. 
According to the AdS/CFT correspondence \cite{adscft}, 
{\em massless} gauge $SU(5)$ fields propagating in 
(infinite) AdS$_5$ is dual to a large-$N$ 
$4D$ conformal field theory (CFT) with a
{\em conserved} $SU(5)$ {\em global} current, $J_{\mu}^{SU(5)}$.
However, if this 
GUT symmetry is spontaneously broken (down to the SM symmetry)
by a scalar vev, $v$, in AdS$_5$,  
then the dual interpretation is that
the CFT matter comes in complete $SU(5)$ multiplets, but
the interactions of the
CFT conserve {\em only} the SM subgroup of the global 
$SU(5)$ symmetry.
Thus, $\partial ^{\mu} J_{\mu}^{X, Y} \equiv \epsilon \neq 0$, 
where
$X$, $Y$ denote the generators of $SU(5)$ not in SM. 
If $v \ll k, \Lambda$ in the AdS$_5$ theory 
(as is the case in the bulk of our model), 
then, in the dual CFT, $\epsilon$ is small.

Adding a Planck brane (the RS model) corresponds to 
putting a UV cut-off of 
$O \left( M_{Pl} \right)$ on the CFT and gauging the {\em full}
$SU(5)$ (global) symmetry of the CFT by $4D$ vector
fields external to the CFT \cite{rs2cft, nima1}. Note that the fact that 
$\partial ^{\mu} J_{\mu}^{X, Y} \neq 0$
means that after coupling to $4D$ gauge fields, we have broken 
$SU(5)$ gauge invariance down to the SM gauge group.
Vacuum polarization effects will then generate masses for 
the $4D$ $X$, $Y$ gauge bosons at the scale
$M_{GUT}$ which is suppressed compared to the Planck scale 
by $\epsilon$. The breaking of gauge invariance looks explicit, 
but as usual this explicit breaking can be 
viewed as Higgs effects in unitary gauge.
The TeV brane corresponds to 
a spontaneous breaking of conformal invariance \cite{nima1, rs1cft}, 
but {\em not} of the
$SU(5)$ symmetry.

The dual interpretation of the low-energy gauge couplings
in Eq.~(\ref{rsgut2}) (along with
Eq.~(\ref{4dgut})) is as follows.
Although $X$, $Y$ gauge bosons have a mass $M_{GUT}$, 
the SM gauge bosons are light.
The splitting of $X$, $Y$ at $M_{GUT}$ from light SM gauge bosons gives rise to 
differential
running of gauge couplings below $\sim M_{GUT}$
with the 
$4D$ $\beta$-function coefficients giving
the $b_i / \left( 2 \pi \right)
\log \left( M_{GUT} / q \right)$ terms in Eq.~(\ref{4dgut}).
The running due to the CFT charged matter 
between the Planck scale (the UV cut-off)
and the TeV scale is {\em mostly} 
(assuming $\epsilon$ is small) $SU(5)$ 
{\em symmetric} and accounts for the (universal)
$\log \big[ O \left( M_4 \right) \big] / \hbox{TeV}$, i.e., $k \pi r_c$ 
enhanced terms in $\alpha_{GUT}^{-1}$, in particular the dominant
$\pi r_c / \alpha_5$ term. 

The $SU(5)$ {\em breaking} part of the {\em running} due to the CFT charged matter 
(which also has a Planckian logarithm)
is suppressed
by $\epsilon$ and corresponds to the $\Delta_i$ in Eq.~(\ref{rsgut2})
-- in particular, the {\em leading}
order in large-$N$ part of this running corresponds to 
contribution of the {\em tree}-level higher-dimensional 
operator, 
whereas,
the {\em sub-}leading in large-$N$ 
part of this running corresponds
to the {\em loop} suppressed non-universal contribution.

Finally, various $O \big[ 1 / \left( 2 \pi \right) \big]$, 
non-log-enhanced terms in Eq.~(\ref{rsgut1}) correspond to 
various
threshold effects in the CFT interpretation.

\subsection{$SU(5)$ breaking on the TeV brane}

The dual interpretation changes when we modify the orbifold
boundary conditions of the minimal module so as to break 
$SU(5)$ on the TeV brane (i.e., $X$, $Y$ vanish on the TeV brane). 
The interpretation now is that 
$SU(5)$ is {\em spontaneously} broken by
the CFT at the 
TeV scale, i.e., the same scale at which 
conformal invariance is broken \cite{contino}, 
on top of the explicit breaking by the CFT dynamics ($\epsilon
\neq 0
$).
Since (on the RS side) the $\mathbf{\bar{5}}_{1,2}$ and $\mathbf{10}_{1,2}$ are
$(+)$ (i.e., do {\em not} vanish) on the
Planck brane, we add fundamental fermions in these representations
in the dual CFT
(these 
will be denoted
by the same symbols)
-- thus, the fundamental fermions
have the same quantum numbers as in the SM, but
are {\em twice} as many. 
These fermions (external to the CFT) couple 
to (fermionic) operators of the CFT (which are in conjugate representations) 
denoted by
${\cal O}_{5_1}$, ${\cal O}_{\bar{10}_1}$ etc.
Since (on the RS side) the quark $SU(2)$ singlet from $\mathbf{\bar{5}}_1$ 
(denoted by $d^c_1$) is $(+, -)$ (i.e.,
vanishes on the
TeV brane) and hence does not have a zero-mode, in the dual CFT,
it must be that the ${\cal O}_{5_1}$ operator of the CFT 
interpolates a {\em massless} composite fermion
of the CFT which marries (i.e., gets a Dirac mass with)
$d^c_1$.
Whereas,
this CFT operator does {\em not} interpolate a massless 
composite to marry the 
fundamental $SU(2)$ doublet lepton from $\mathbf{\bar{5}}_1$ (denoted by
$L_1$). The fact that massless composite 
fermions of the CFT below TeV do {\em not} come in
complete $SU(5)$ multiplets is 
consistent because $SU(5)$ is broken (spontaneously) by 
the CFT at that scale. 
$L_1$
can mix with
the CFT composites -- 
the resultant massless state corresponds to the SM $SU(2)$ doublet lepton
(denoted by $L$).
The degree of this mixing depends on the anomalous dimension
of the fermionic CFT operator, ${\cal O}_{5_1}$ 
which is related on the RS side
to the 
fermion mass parameter $c$.
A similar analysis can be done for the other SM fermions. 

The coupling of fundamental fermions to fermionic
CFT operators is essential to generating Yukawa couplings since the
Higgs doublet is a 
composite of the CFT. In this way, the RS scenario with bulk fermions 
realizes the idea proposed in reference \cite{kaplan}.

The contribution of the fundamental fermions to running of the gauge couplings
(above TeV scale) is universal (for $i = 1, 2, 3$)
since they are in complete $SU(5)$ multiplets (for
example, $L_1$ and $d^c_1$) -- the modification
of this running due to mixing of 
these fermions with CFT composites is also universal since
$L_1$ and $d^c_1$ couple to the {\em same} CFT operator, ${\cal O}_{5_1}$
and hence the mixing is $SU(5)$ symmetric.

\subsubsection{Proton decay}
\label{cfttevproton}

The $X$, $Y$ gauge bosons (either the fundamental 
ones with a mass of $M_{GUT}$ or the bound states of the CFT with
$X$, $Y$ quantum numbers and with masses
$\sim$ TeV) do {\em not} couple 
SM fermion $L$
to $d^c$
since these fermions 
(which are
mixtures of fundamental and CFT fields) have their origin in
{\em different} $\mathbf{\bar{5}}$ fundamental fields. 
So, 
proton decay from exchange of $X$, $Y$ gauge bosons is absent. 

As mentioned earlier,
to forbid proton decay from higher-dimensional operators, on the RS side, 
we introduce  
a bulk $\tilde{U}(1)$ gauge symmetry which is a linear combination of
$U(1)_B$ and $U(1)_Y$. 
Spectators on TeV brane cancel all $\tilde{U}(1)$ 
anomalies. 
The $\tilde{U} (1)$ gauge symmetry is broken 
by the Planckian vev of a SM singlet scalar living
on the Planck brane. The dual picture is 
that the CFT and the fundamental fermions coupled to it
have an exact $\tilde{U} (1)$ 
symmetry which is gauged by a $4D$ vector field. It is {\em not}
that the $4D$ gauging of the $\tilde{U} (1)$ symmetry 
is protecting the theory from 
excessive proton decay -- in fact, in the dual picture, the
$4D$ $\tilde{U} (1)$ gauge theory is Higgsed near the Planck 
scale and at this scale operators violating $\tilde{U} (1)$ and $U(1)_B$ are 
allowed. The central point is that all such violations of $U(1)_B$ 
are strongly irrelevant in the IR of the CFT coupled to fundamental 
fermions and light gauge fields.\footnote{Of course, $\tilde{U}(1)$ gauge
symmetry is also 
spontaneously broken by the Higgs vev, 
but this breaking still preserves
baryon-number. Hence higher-dimensional operators, 
generated by spontaneous breaking of scale invariance at $\sim$ TeV
and suppressed
by that scale (which are dual to
TeV brane-localized operators on the RS side), 
will preserve baryon-number even though
they might violate $\tilde{U}(1)$ {\em after} electroweak symmetry breaking.
Recall that SM fermions might have an
admixture of CFT composites so that
these operators can give interactions
between SM fermions and so we want them to preserve baryon-number.}
In other words, at sub-Planckian energies, $U(1)_B$ is an accidental, 
global symmetry very much as 
in the SM.\footnote{The mild distinction is that, in the 
SM, baryon-number is an accidental {\em classical} symmetry, but is quantum
mechanically {\em anomalous} leading to non-perturbative violation.}
This is the dual of the fact that 
$U(1)_B$ is unbroken on the RS side throughout the bulk
and on the TeV brane.
%

In the dual picture, 
the spectators are composites of the
CFT and have Yukawa
couplings to the Higgs (which itself is a CFT composite) which give them 
electroweak scale masses when the Higgs acquires a vev. 
It is interesting to see how cancellation of anomalies
involving $\tilde{U} (1)$ 
works in the {\em UV}
of the dual theory -- 
in the IR, these anomalies cancel between
SM fermions and spectators just as on the RS side. 
The CFT generates two kinds of {\em massless} composite
fermions: (a) the spectator fermions and (b) 
fermions 
required to marry (the fundamental fermions corresponding to) the $(+, -)$ 
states in the $\mathbf{\bar{5}}$'s and $\mathbf{10}$'s
(i.e., $d^c_1$, $L_2$ etc.)
as mentioned above -- the SM and
$\tilde{U} (1)$ quantum numbers of these
composites are opposite to 
those of the states in $\mathbf{\bar{5}}$'s and $\mathbf{10}$'s
they marry.
One can check that these composite fermions do {\em not} have SM, 
(SM)$^2 \times \tilde{U} (1)$ 
and 
SM$\times \big[ \tilde{U} (1) \big]^2$ anomalies, but {\em do} have a non-zero
$\big[ \tilde{U} (1) \big] ^3$ anomaly. By the 't Hooft anomaly matching
condition, the anomalies of
the CFT {\em by itself}
in the {\em UV} must be the {\em same} as that of its 
{\em massless} (fermionic) composites. It can be checked that, in the UV,
this $\big[ \tilde{U} (1) \big] ^3$ anomaly of the
CFT is canceled by that of the {\em fundamental} 
$\mathbf{\bar{5}}$'s and
$\mathbf{10}$'s -- the latter
also do {\em not} have SM, (SM)$^2 \times \tilde{U} (1)$ 
and SM$\times \big[ \tilde{U} (1) \big]^2$ anomalies.
Thus, the CFT {\em and} external $\mathbf{\bar{5}}$'s and
$\mathbf{10}$'s together do not have any anomalies in the
UV and hence both $SU(5)$ and 
$\tilde{U} (1)$ can be gauged.

The fact that 
the Higgs doublet is {\em not} part of a complete $SU(5)$ multiplet
again is consistent with the fact that $SU(5)$ is spontaneously broken
by the CFT at the TeV scale and 
the Higgs doublet is a composite produced {\em below} 
that
scale. 

\subsection{$SU(5)$ breaking on the Planck brane}

If $SU(5)$ is broken by an orbifold boundary condition 
at the Planck brane,
(i.e.,
the $X$, $Y$ vanish on the Planck brane),
then the dual interpretation is that {\em only} the SM subgroup of 
the 
(approximate, up to $\partial ^{\mu} J_{\mu} ^{X, Y} = \epsilon \neq 0$)
$SU(5)$ global symmetry of the CFT is gauged \cite{contino, gns}. 
Thus,
there
are {\em no} $X$, $Y$ gauge bosons with mass $\sim M_{GUT}$ (unlike section
\ref{cftminimal}) so that
the SM gauge bosons cause differential running below the UV cut-off
$\sim O \left( M_{Pl} \right)$, 
rather than $M_{GUT}$, i.e., $b_i \log \left( M_{GUT} / M_Z \right) \rightarrow
b_i \log \big[ O \left( M_{Pl} \right) / M_Z 
\big]$ in Eq.~(\ref{rsgut1}). Since $M_{GUT} \sim O \left( M_{Pl} \right)$,
the large logarithms are unchanged. 
It is amusing that on the RS side, 
the gauge theory is $SU(5)$, whereas in the dual CFT
it is the SM gauge group.
When conformal invariance
is broken at $\sim$ TeV, (global) 
$SU(5)$ is still preserved (up to, of course, $\epsilon$). 
Since (on the RS side)  
$d^c_1$
is $(-, +)$
(i.e., vanishes on the Planck
brane), we add 
a fundamental $L_1$ (and {\em not} $d^c_1$)
and couple it to (the $SU(2)$ doublet part of) the
operator 
${\cal O}_{5_1}$.
Similarly, we add 
$d^c_2$
and 
couple it to 
the ({\em different}) operator
${\cal O}_{5_2}$. Since the {\em full}
$SU(5)$ symmetry is {\em not} gauged, 
the fundamental fermions (external to the CFT)
do {\em not} have to be in complete $SU(5)$ multiplets, but they do 
have to be {\em parts} of $SU(5)$ multiplets since they have to couple
to CFT operators which are in (complete) $SU(5)$ multiplets \cite{gns}. Thus,
the understanding of quantum numbers of the SM fermions is intact even
though we do {\em not} have $SU(5)$ {\em gauge} invariance in the dual CFT theory 
\cite{gns}.   
Thus, the set of fundamental fermions 
is the {\em same}
as the set of SM fermions -- 
in fact, up to mixing with the CFT composites as before
(due to their coupling to the corresponding operators),
these {\em are} the SM quarks and leptons (i.e., they are 
dual to the fermion zero-modes of RS side).
As before, bound states of the CFT with $X$, $Y$ quantum
numbers can{\em not} couple these massless fermions to each other. 

If the mixing of fundamental fermions with the CFT composites is small
(dual to $c \stackrel{>}{\sim} 1/2$), then the contribution of
these fermions to running
of gauge couplings
is universal since they can be ``assembled'' into
complete $SU(5)$ multiplets
(for example,
$L_1$ and $d^c_2$). However, this mixing (or dressing
by CFT interactions) can be large
(dual to $c \stackrel{<}{\sim} 1/2$) and {\em different} for $L_1$ and $d^c_2$
since they couple to CFT operators, ${\cal O}_{5_1}$ and ${\cal O}_{5_2}$,
with
{\em different} anomalous dimensions. 
Since the contribution to running is affected by this CFT dressing, it 
is not universal in this case, even though (as in
the case of small mixing) $L_1$
and $d^c_2$ are equivalent to a complete $SU(5)$ multiplet. Of course,
if ${\cal O}_{5_1}$ and ${\cal O}_{5_2}$ have similar anomalous dimensions,
then running is again universal. 

As in section \ref{cfttevproton}, baryon-number is an accidental, global symmetry
of the CFT and fundamental fermions. 
This symmetry is gauged, but again the gauge symmetry is 
broken at Planckian scales and offers no protection at low energies. Unlike
in section \ref{cfttevproton}, 
spectators are fundamental fields external to the CFT
which also get 
Planckian masses at the same time as the $\tilde{U} (1)$ gauge field.
Thus, the protective $U(1)_B$ accidental symmetry is quantum mechanically
anomalous. The anomaly in baryon-number is due to the 
fundamental fermions, the CFT itself
does  not contribute to the anomaly since it does {\em not} 
generate any {\em massless} composite fermions.
Thus, proton decay is suppressed just as in the SM by an accidental global {\em 
anomalous}
$U(1)_B$ symmetry.

Finally, the dual interpretation of 
the Higgs triplet on the TeV brane is that such
a composite of the CFT 
{\em is}
expected from the (approximate) $SU(5)$ (global) symmetry of the CFT sector 
(which
survives even after 
conformal invariance is broken at TeV). The Higgs triplet 
can{\em not} mediate proton decay because of the accidental baryon-number
symmetry mentioned above. 
%

Another issue concerning the Higgs triplet is why it does not acquire a 
phenomenologically
dangerous vev. Indeed,
if $\epsilon = 0$ (i.e., the
CFT has {\em exact} $SU(5)$ global symmetry) and we consider the CFT 
in isolation, 
then a vev for the Higgs doublet implies a vev for the Higgs triplet
(again, both are CFT composites).
However, 
due to the small ($O (\epsilon)$) explicit breaking of the $SU(5)$ (global)
symmetry of the CFT,  
the mass of Higgs triplet can be different 
than that of the Higgs doublet (this is the
dual of $\Sigma$ vev splitting the Higgs doublet from the triplet) so 
that the Higgs triplet
need not acquire a vev. 
Also, the 
CFT is {\em not} isolated, but is 
coupled to $SU(5)$ non-symmetric fundamental fields. 
So, 
{\em loop} corrections 
can further split
the Higgs triplet from the doublet (just as on the RS side).

\section{Conclusions}
\label{conclude}

The simplest way to summarize the connection between unification and  the 
RS solution to the hierarchy problem is to 
adopt the CFT dual viewpoint, in the sense of the AdS/CFT correspondence. 
From this viewpoint, the RS higher-dimensional effective field theory studied 
here is a very useful, weakly coupled dual description of a strongly coupled $4D$ 
Higgs sector, KK excitations being dual to some of the strongly bound 
composites. The Higgs boson is an infrared composite of this sector, so that
the hierarchy problem is solved by compositeness, without SUSY. 
SM gauge bosons and fermions are fundamental fields coupled to the strong 
Higgs sector. A key to having an extra-dimensional description which is weakly 
coupled throughout the extra dimension is that the dual $4D$ description 
should be strongly coupled throughout the large Planck-weak hierarchy. This is 
achieved by having the strongly coupled sector be approximately conformal
over the large hierarchy. 

Let us now turn to unification.
In $4D$ supersymmetric GUTs, MSSM two-loop running 
leads to gauge coupling unification to within $\sim 4 \%$. The discrepancy 
can naturally be ascribed to threshold corrections (although there 
is no predictivity at this level, due to free parameters of the GUT). 
Only slightly less striking is $4D$ {\em non}-supersymmetric GUTs, where SM
running alone leads to gauge coupling unification to within $\sim 20 \%$.
However, this discrepancy is now too big to be naturally ascribed to GUT 
threshold corrections. Of course, we can add new particles below the GUT scale 
which alter the differential running by the small amount needed to achieve 
unification. This is what happens when we supersymmetrize the SM in 
order to solve the hierarchy problem. But if the new particles we are adding 
to solve the hierarchy problem are strongly coupled over the hierarchy as in
the RS scenario, 
there is a second 
attractive possibility for unification. We can take the particles of the strong 
sector to come in {\em complete} 
$SU(5)$ multiplets but with interactions which are 
only {\em approximately} $SU(5)$ symmetric. The SM subgroup of $SU(5)$ is 
however an exact symmetry of the strong sector and is gauged.
Thus the strong sector will naturally 
make small contributions to {\em differential} running because the
{\em strong} interactions which
{\em non-negligibly} dress contributions to gauge coupling running
are {\em weakly} 
$SU(5)$-violating. The small 
discrepancies in SM unification can be easily ascribed to these 
small corrections from the strong sector to differential running. 

It is 
interesting that from the dual 
$4D$ viewpoint the small corrections are corrections to 
{\it running} and therefore logarithmically enhanced, while in the $5D$ RS 
viewpoint, the small corrections are threshold type corrections which are 
volume-enhanced. Either way these 
corrections needed for unification are {\em not} 
calculable without knowing details of the GUT and Planck scales, 
which is {\em qualitatively} similar to the situation in supersymmetric GUTs.
Our central result is that we are able to maintain the SM level of 
unification in a {\em calculable} way in a theory of a composite Higgs
and to point out a naturally small source 
of corrections which ``fix'' SM running with respect to unification. 
These are very difficult problems to address within
traditional approaches to strongly interacting Higgs sectors.

From the dual $4D$ viewpoint, the central object which is unified is the 
conformal Higgs sector, in that it has an approximate global $SU(5)$ symmetry.
In order for fundamental SM fermions to get non-negligible Yukawa 
couplings, they must couple to fermionic operators of the conformal Higgs sector. 
Since these operators come in $SU(5)$ multiplets, in order to preserve SM 
gauge invariance 
the SM fermions must have gauge quantum numbers which can be 
embedded in $SU(5)$, providing a partial 
explanation of the quantum numbers we observe. 
Excessive proton decay is avoided if the Higgs sector 
coupled to fundamental fermions has an accidental baryon number symmetry (which 
might be anomalous upon SM gauging, as in the SM). 

The near meeting of gauge couplings at high 
energies and the pattern of fermion quantum
numbers within the SM provide tantalizing hints of unification. 
This is usually taken as 
evidence in favor of SUSY where gauge coupling unification is
{\em quantitatively} improved while solving 
the hierarchy problem. However, we have shown
here that there is an 
attractive option for unification within the RS scenario
for solving the hierarchy problem.

\section*{Acknowledgments}

We thank Markus Luty, John March-Russell, Alex Pomarol and Neal Weiner for
discussions. The work of
K.~A.~was supported by the Leon Madansky fellowship and by
NSF Grant P420D3620414350. The work of A.~D.~was supported by NSF 
Grants P420D3620414350 
and
P420D3620434350. The work of 
R.~S.~was supported in part by
a DOE Outstanding Junior Investigator Grant P442D3620444350 and by 
NSF Grant P420D3620434350.

\end{document}